\documentclass[conference]{IEEEtran}
\ifCLASSINFOpdf
\else
\fi

\usepackage{graphicx}
\usepackage{amsmath}
\usepackage{amsfonts}

\begin{document}
%
\title{Optimal time sharing in underlay cognitive radio systems with RF energy harvesting}

\author{\IEEEauthorblockN{Valentin Rakovic, Daniel Denkovski, Zoran Hadzi-Velkov and Liljana Gavrilovska}
\IEEEauthorblockA{Ss. Cyril and Methodius University in Skopje\\
Faculty of Electrical Engineering and Information Technologies\\
Skopje, R. Macedonia\\
Email: \{valentin, danield, zoranhv, liljana\}@feit.ukim.edu.mk}}


\maketitle

\begin{abstract}
Due to the fundamental tradeoffs, achieving spectrum efficiency and energy efficiency are two contending design challenges for the future wireless networks. However, applying radio-frequency (RF) energy harvesting (EH) in a cognitive radio system could potentially circumvent this tradeoff, resulting in a secondary system with limitless power supply and meaningful achievable information rates. This paper proposes an online solution for the optimal time allocation (time sharing) between the EH phase and the information transmission (IT) phase in an underlay cognitive radio system, which harvests the RF energy originating from the primary system. The proposed online solution maximizes the average achievable rate of the cognitive radio system, subject to the $\varepsilon$-percentile protection criteria for the primary system. The optimal time sharing achieves significant gains compared to equal time allocation between the EH and IT phases.
\end{abstract}



%
\IEEEpeerreviewmaketitle

\section{Introduction}

Energy harvesting (EH) is a highly promising technology which can support continuous power supply to energy-constrained communication system, such as machine-to-machine and sensor networks. Usually the main goal of the EH communication systems is to maximize the system throughput (i.e., the average achievable rate) under some causality constraints \cite{cite1,cite2,cite3}. The harvested energy can originate from the far-field radio frequency (RF) radiation of another device (e.g., a closely located communications transmitter which can transfer power and information simultaneously), or from the ambient RF radiation. For example, \cite{cite4} shows that, if a mobile phone transmits with power of 0.5W, an EH device at distances of 1, 5, and 10 meters can harvest as much as 40mW, 1.6mW, and 0.4mW, respectively. If a source is used for simultaneous information and power transfer, there exists a fundamental tradeoff between the amount of transferred energy and the achievable information rates (known as the rate-energy tradeoff) due to the infeasibility of the receiver to extract information and energy simultaneously from the same signal \cite{cite5}.

If an energy-constrained system does not have dedicated communication resources (e.g., dedicated bandwidth), it can still be implemented as a secondary user (SU) system in a cognitive radio (CR) setup \cite{cite6}. The CR system utilizes one of the following three methods for spectrum sharing: interweave (i.e. opportunistic), underlay and overlay spectrum sharing \cite{cite7}. In the interweave approach the CR devices perform spectrum sensing to detect the unoccupied incumbent bands and then access the channel in this band. In the underlay approach, the CR devices operate in the same band as the primary users (PUs), and are allowed to interfere with the primary system under some predefined threshold (e.g., based upon the criterion for the $\varepsilon$-percentile protection of the primary system \cite{cite8}, \cite{cite9}). The overlay sharing approach differs from the opportunistic and underlay approaches, because the former approach envisions cooperation between the PU and SU systems. In the overlay CR systems, both the PU and SU systems adapt their transmissions in order to accommodate each other at the expense of increased complexity, whereas, in the case of the opportunistic and underlay spectrum sensing, the primary system is unaware of the secondary one. Actually, the underlay spectrum sharing is the simplest method and the easiest to implement in practice.

In this paper, we study the underlay CR systems with RF energy harvesting capabilities at the secondary transmitter. Using the practically limitless but free power supply from the far-field RF radiation, the secondary system can maintain continuous information flow with meaningful achievable information rates. More specifically, we focus on the scenario where the SU and PU systems co-exist in the same communications band, and additionally the PU transmitter, during its normal communication with the PU receiver, also serves as a source of RF energy to the SU transmitter. Actually, the secondary transmitter exploits the signal coming from the PU system, which is considered to be harmful interference in the non-EH secondary systems. We aim at maximizing the performance of the RF energy harvesting, subject to the $\varepsilon$-percentile protection criterion of the primary system. To the best of the authors' knowledge, this scenario has not been considered in the literature. The paper proposes an online solution for the optimal time resource allocation between the EH and information transmission (IT) phases in the SU system. The proposed online solution is designed to maximize the SU average achievable rate while avoiding interfering with the PU system but requires the instantaneous CSI of several fading channels of the primary and secondary systems at the SU transmitter.

The reminder of the paper is structured as follows. Section II presents the recent related works in the area. Section III elaborates the proposed and adopted CR system mode. Section IV presents the proposed online solution. Section V focuses on the performance analysis the proposed online solution. Section VI concludes the paper.

\section{Related Work}
Most of the papers that combine CR and EH are focusing on scenarios where the CR systems perform energy harvesting from outside sources other than RF sources or energy harvesting from ambient RF radiation. So far, none of the exiting works have studied the scenario where the CR systems in underlay sharing mode harvest the RF energy originating from the primary system.

Moreover, the majority of the related works focus on the case of opportunistic spectrum access, where the CR system performs the RF energy harvesting and information transmission on separate channels (frequency bands). In \cite{cite10}, the authors investigate SU devices that can switch between the opportunistic spectrum access and RF energy harvesting, and develop an optimal mode selection policy that maximizes the averaged SU system rate by balancing between the instantaneous rate and the amount of harvested RF energy. The work in \cite{cite11} focuses on an opportunistic SU system that performs energy harvesting from outside sources (but not RF sources). More specifically, \cite{cite11} determines both the sensing duration and sensing threshold that maximize the average rate of the SU system for a given amount of harvested energy. The paper \cite{cite12} studies a channel selection policy in a multi-channel opportunistic CR system, in which the SUs can select a specific frequency band either for data transmission or for RF energy harvesting from occupied channels. The papers \cite{cite13} and \cite{cite14} extend the system model from \cite{cite12} by introducing incomplete CSI at the SUs. In \cite{cite13} the optimal channel selection policy is obtained by considering errors in the spectrum sensing process. The authors in \cite{cite14} propose an online learning algorithm that obtains cognitive radio related environmental parameters, like channel availability, interference level etc. The proposed algorithm is based on the Markov decision process and uses the environmental observations for adjusting the channel selection, thus maximizing the secondary system's throughput.

In \cite{cite15} and \cite{cite16}, the authors extend the system model of the former papers by introducing stochastic geometric approaches for performance assessment of CR systems with opportunistic spectrum sharing and RF energy harvesting. The paper \cite{cite15} considers a scenario where both the RF-powered SUs and the PUs follow independent homogeneous Poisson Point Processes (PPP) and communicate with their intended receivers at fixed distances. Moreover, it is assumed that each PU device randomly accesses the spectrum with a given probability and that each transmitting PU is centered at a guard zone and a harvesting zone that is inside the guard zone. Each SU transmitter either harvests the energy of a nearby PU transmitter if it lies in its harvesting zone, transmits if it is outside the guard zones of the PU transmitters, or it stays idle otherwise. The authors derive the optimal transmit power and SU density that maximize the secondary system's rate, subject to the outage constraints of both PU and SU systems. In \cite{cite16}, the authors investigate a CR based device-to-device (D2D) communication capable of harvesting the ambient interference resulting from the concurrent macro base station transmissions. After harvesting sufficient energy, the CR transmitters opportunistically access a predefined nonexclusive channel that satisfies the interference constraints of the CR and macro cell communication.

In the case of overlay spectrum sharing, \cite{cite17} proposes a novel approach applicable only to overlay systems, denoted as the energy-information cooperation. The PU network provides both spectrum and energy to the SU network so that the SU network will aid (in a relaying fashion) the PU transmission in return. Assuming availability of non-causal PU information, the proposed energy-information cooperation can achieve substantial performance gain compared to the conventional information cooperation only. However, the system model assumptions, regarding the full PU-SU cooperation and non-causal information at the SUs limit the practical applicability of the approach.

An underlay CR system with RF energy harvesting has not been studied so far.

\section{System model}

\begin{figure}[!t]
\centering
\includegraphics [width=2.5in]{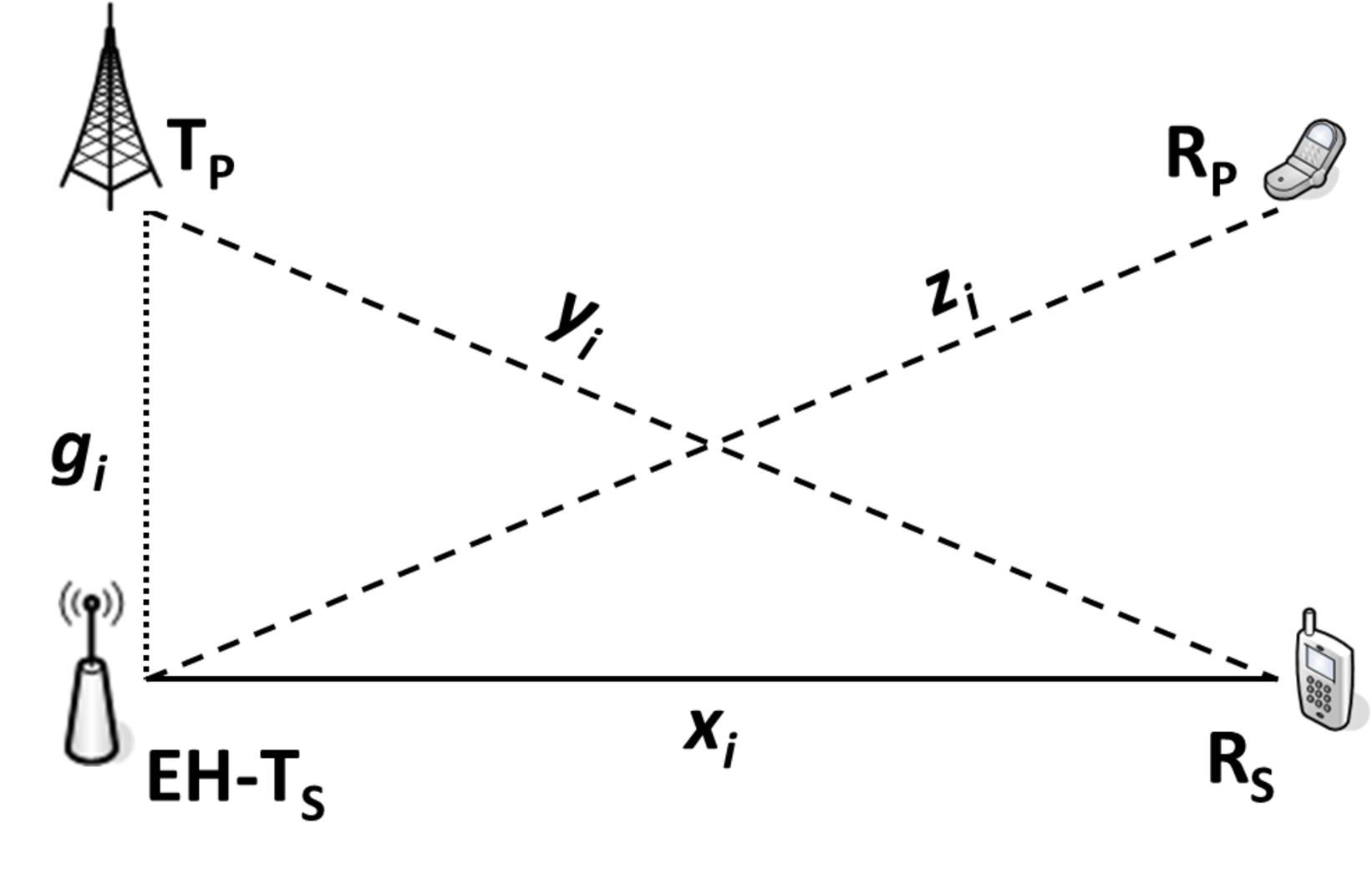}
\caption{System model}
\label{fig:1}
\end{figure}

In our considered underlay CR system, the SU is a point-to-point communication system, and consists of an EH transmitter ($EH-T_S$) and a secondary receiver ($R_S$). The PU is also a point-to-point communication system, and consists of a primary transmitter ($T_P$) and a primary receiver ($R_P$). The $EH-T_S$ is equipped with a rechargeable battery, which is recharged by harvesting the RF energy in the common communication channel (i.e., the common frequency band). Without any loss of generality, the $EH-T_S$ battery capacity is assumed to be large enough to accumulate the entire amount of radiated PU energy per communication slot. For example, the radiated power from a DVB-T transmitter with $P_T=10kW$ at a distance of 100m is approximately 1.1mW (assuming Free Space Path Loss). This is significantly lower than the (power) capacity of conventional mobile device batteries that range in the magnitude of several Watts. Apart from being an information source for the $R_P$, and an RF energy source for the $EH-T_S$, the RF energy radiated from the $T_P$ is also a cause of interference for the $R_S$. The considered cognitive system is depicted in Fig.~\ref{fig:1}.

It is assumed that the system operates in a block fading environment, where the channel gains remain constant during one channel block of duration $T$, but change from one block to the other. The time in the secondary system is divided into slots of equal duration, and each time slot lasts exactly one channel block of duration $T$.

The total number of considered times slots is $M$. In the $i^{th}$ time slot, we define the following fading power gains (depicted in Fig.~\ref{fig:1}): the gain of the $EH-T_S$ to $R_S$ channel is denoted by $x_i$, the gain of the $T_P$ to $R_S$ channel by $y_i$, the gain of the $T_P$ to $EH-T_S$ by $g_i$, and the gain of the $EH-T_S$ to $R_P$ by $z_i$. The power of the additive Gaussian noise at the $R_S$ is $\sigma^2$. We assume that the $T_P$ continuously transmits with fixed output power $P_T$.

During each time slot, the secondary system operation is divided in two phases: an EH phase and an information transmission (IT) phase. In the EH phase, $EH-T_S$ harvests the RF radiation from $T_P$, whereas it transmits during the IT phase. However, $EH-T_S$ adapts the duration of the phases from one slot to the other. In $i^{th}$ time slot, the duration of the EH phase is $(1-\alpha_i)T$, and the duration of the subsequent IT phase is $\alpha_iT$, where $\alpha_i$ is the time-sharing parameter. As seen in the following section, the time-sharing parameter $\alpha_i$ ($0 \le \alpha_i \le 1$) is actually our optimization variable.

In the rest of the paper, we assume a normalized unit block time ($T = 1$) without loss of generality, so we use the terms of energy and power interchangeably. Note, we assume that during the IT phase, the $EH-T_S$ completely drains the energy harvested in its battery during the preceding EH phase.

In the beginning of each time slot, the secondary system is assumed to have the instantaneous CSI available for the channels: $x_i$, $y_i$, $g_i$, $z_i$ i.e. it knows the referred channel power gains before the SU's transmission. In practice the channel gain $x_i$ can be estimated prior to the transmission, either via a training sequence or in the uplink period if a time division duplex (TDD) scheme is used (assuming channel reciprocity). Due to the constant and known PU transmit power $P_T$ the channel gain $g_i$ can be estimated by the $EH-T_S$, prior to the secondary communication, exploiting the primary system training sequence. The channel gain $y_i$ can be estimated by the $R_S$ (in the same fashion as for channel gain $g_i$) and reported to the $EH-T_S$. Finally, the channel gain $z_i$, being the most complex one to estimate and fed back, can be either obtained from a possible PU-SU cooperation, or in the PU uplink period if assuming a TDD scheme for the PU system.

\subsection{Primary system constraints}
We adopt $\varepsilon$-percentile protection criteria for the primary system \cite{cite8},\cite{cite9}, which means that the outage probability of the primary system caused by the interference from the $EH-T_S$ transmission does not exceed the $\varepsilon$ threshold:

\begin{equation}\label{eq:1}
\Pr \left\{ {{P_i}{z_i} \ge {\gamma _{th}}} \right\} \le \varepsilon,
\end{equation}

\noindent where $P_i$ represents the transmit power of the $EH-T_S$  in time slot $i$, and $\gamma_{th}$ is the maximum tolerable interference power from $EH-T_S$ at the $R_P$ (denoted as the PUI threshold).

Assuming ideal harvesting and conversion efficiencies, the energy harvested by $EH-T_S$ in the EH phase of slot $i$ is given by $E_i = (1-\alpha_i)g_iP_T$. Consequently, average transmit power from the $EH-T_S$ during the IT phase of slot $i$ is given by $P_i = E_i/\alpha_i = (1-\alpha_i)g_iP_T/(\alpha_i)$. Hence, (\ref{eq:1}) is rewritten as:

\begin{equation}\label{eq:2}
\Pr \left\{ {\frac{{{\rm{1}} - {\alpha _i}}}{{{\alpha _i}}}{g_i}{P_T}{z_i} \ge {\gamma _{th}}} \right\} = \Pr \left\{ {{\alpha _i} \le \frac{{{g_i}{P_T}{z_i}}}{{{\gamma _{th}} + {g_i}{P_T}{z_i}}}} \right\} \le \varepsilon.
\end{equation}

In order to express (\ref{eq:2}) in terms of a statistical or a time average, we introduce the indicator function as follows:

\begin{equation}\label{eq:3}
I({\alpha _i}) = \left\{ {\begin{array}{*{20}{c}}
{0,}&{{\alpha _i} > \frac{{{g_i}{P_T}{z_i}}}{{{\gamma _{th}} + {g_i}{P_T}{z_i}}}}\\
{1,}&{\text{otherwise}}
\end{array}} \right.
\end{equation}

Now, (\ref{eq:2}) can be rewritten as $\mathbb{E}\left[ {I({\alpha _i})} \right] \le \varepsilon$ , where $\mathbb{E}\left[  \cdot  \right]$  denotes the expectation.

\section{Optimal time resource allocation}

Given the available transmit power $P_i$ in the beginning of IT phase in slot $i$, the achievable rate of the secondary system during the IT phase is given by \cite{cite18}:

\begin{equation}\label{eq:4}
\begin{array}{c}
{R_i} = {\alpha _i}{\log _2}\left( {1 + \frac{{1 - {\alpha _i}}}{{{\alpha _i}}}\frac{{{g_i}{P_T}{x_i}}}{{{y_i}{P_T} + {\sigma ^2}}}} \right)\\
 = {\alpha _i}{\log _2}\left( {1 + \frac{{1 - {\alpha _i}}}{{{\alpha _i}}}{S_i}} \right),
\end{array}
\end{equation}

\noindent where $S_i$ is expressed as:

\begin{equation}\label{eq:5}
{S_i} = \frac{{{g_i}{P_T}{x_i}}}{{{y_i}{P_T} + {\sigma ^2}}}.
\end{equation}

The rate (\ref{eq:4}) can be achieved by Gaussian distributed codewords transmitted by $EH-T_S$ and reliably decoded by $R_S$ at the end of slot $i$.

We now define the following optimization problem (\textbf{P1}):

\begin{equation}\label{eq:6}
\begin{array}{c}
\mathop {\max }\limits_{{\alpha _i},\forall i} \bar R\\[1mm]
s.t.\Pr \left\{ {{P_i}{z_i} \ge {\gamma _{th}}} \right\} \le \varepsilon,\\[1mm]
{\rm{     }}0 < {\alpha _i} < 1,
\end{array}
\end{equation}

\noindent where  $\bar R = \mathbb{E}\left[ R \right]$ is the average achievable rate (i.e., expectation over all fading channel states). For sufficiently high number of time slots $M$, such that $M \to \infty $, \textbf{P1} can be equivalently written as (\textbf{P2}):

\begin{equation}\label{eq:7}
\begin{array}{c}
\mathop {\max }\limits_{{\alpha _i},\forall i} \frac{1}{M}\sum\limits_{i = 1}^M {{R_i}} \\
s.t.{\rm{ }}\frac{1}{M}\sum\limits_{i = 1}^M {I\left( {{\alpha _i}} \right)}  \le \varepsilon ,\\
0 < {\alpha _i} < 1,
\end{array}
\end{equation}

\noindent where $R_i$ is given by (\ref{eq:4}). Note, the constraint in (\ref{eq:7}) is the PU protection criterion expressed in terms of the indicator function (\ref{eq:3}). We now apply the Lagrange duality method \cite{cite8}, then introduce the Lagrange multiplier $\lambda$ associated with the constraint in (\ref{eq:7}), and then decouple (\textbf{P2}) into parallel sub-problems (one for each fading state), which yields:

\begin{equation}\label{eq:8}
\mathop {\max }\limits_{{0 < \alpha _i < 1}} \left( {{R_i} - \lambda I\left( {{\alpha _i}} \right)} \right).
\end{equation}

\subsection{Optimal solution}

The optimization problem \textbf{P2} is a convex optimization problem, because the objective function is a concave function ($x\log \left( {1 + {1 \mathord{\left/  {\vphantom {1 x}} \right.  \kern-\nulldelimiterspace} x}} \right)$ is concave), whereas the constraint is a step function \cite{cite19}. In the cases when $I(\alpha_i)=0$, the optimal $\alpha_i$ that maximizes the achievable rate for a particular fading state $i$, (\ref{eq:8}) is solved similarly as [18, Eqs. (38) and (40)], yielding:

\begin{equation}\label{eq:9}
{\alpha _{i1}} = \frac{{{S_i}}}{{{S_i} + {z_0} - 1}},
\end{equation}

\noindent where $z_0>1$ is the unique solution to the equation:

\begin{equation}\label{eq:10}
z\ln z - z - {S_i}{\rm{  +  1  =  0}}{\rm{.}}
\end{equation}

However, if the primary link outage constraint is violated, which occurs when $\alpha_{i1} \in [0, g_iP_Tz_i/(g_iP_Tz_i +\gamma_{th})]$, the optimal solution must guarantee that PU outages do not occur while the rate is as high as possible (Fig. 3 in [18]), thus:

\begin{equation}\label{eq:11}
{\alpha _{i2}} = \frac{{{g_i}{P_T}{z_i}}}{{{g_i}{P_T}{z_i} + {\gamma _{th}}}}.
\end{equation}

Since the outage probability threshold in the PU system is set to be $\varepsilon$, we can tolerate outages in $100\cdot\varepsilon$ percent of the time in order to maximize the long-term average achievable rate (i.e., over $M \to \infty$ time slots). This threshold can be achieved by the appropriate selection of the Lagrange multiplies $\lambda$ that appears in (\ref{eq:8}). It can be determined numerically such that the constraint in (\ref{eq:7}) is satisfied. In a practical system, $\lambda$ can be computed during the system training phase and then it can be used for online determination of the time-sharing parameter $\alpha_i$, as according to the solution of (\ref{eq:8}), which is given by:

\begin{equation}\label{eq:12}
\alpha _i^* = \left\{ {\begin{array}{*{20}{c}}
{{\alpha _{i1}},}&{I\left( {{\alpha _{i1}}} \right) = 0}\\
{{\alpha _{i1}},}&{I\left( {{\alpha _{i1}}} \right) = 1 \quad \textrm{and} \quad R_{i1}^{} > {R_{i2}}}\\
{{\alpha _{i2}}}&{I\left( {{\alpha _{i1}}} \right) = 1 \quad \textrm{and} \quad R_{i1}^{} \le {R_{i2}}}
\end{array}} \right.,
\end{equation}

where $R_{i1}$ and $R_{i2}$ are defined as:

\begin{equation}\label{eq:13}
\begin{array}{c}
R_{i1}^{} = f(\alpha _{i1}^{}) - \lambda I(\alpha _{i1}^{}),\\ [1mm]
{R_{i2}} = f({\alpha _{i2}}),\\ [1mm]
f(\alpha ) = \alpha {\log _2}\left( {1 + \frac{{1 - \alpha }}{\alpha }S} \right).
\end{array}
\end{equation}

\begin{figure}[!t]
\centering
\includegraphics [width=3.2in]{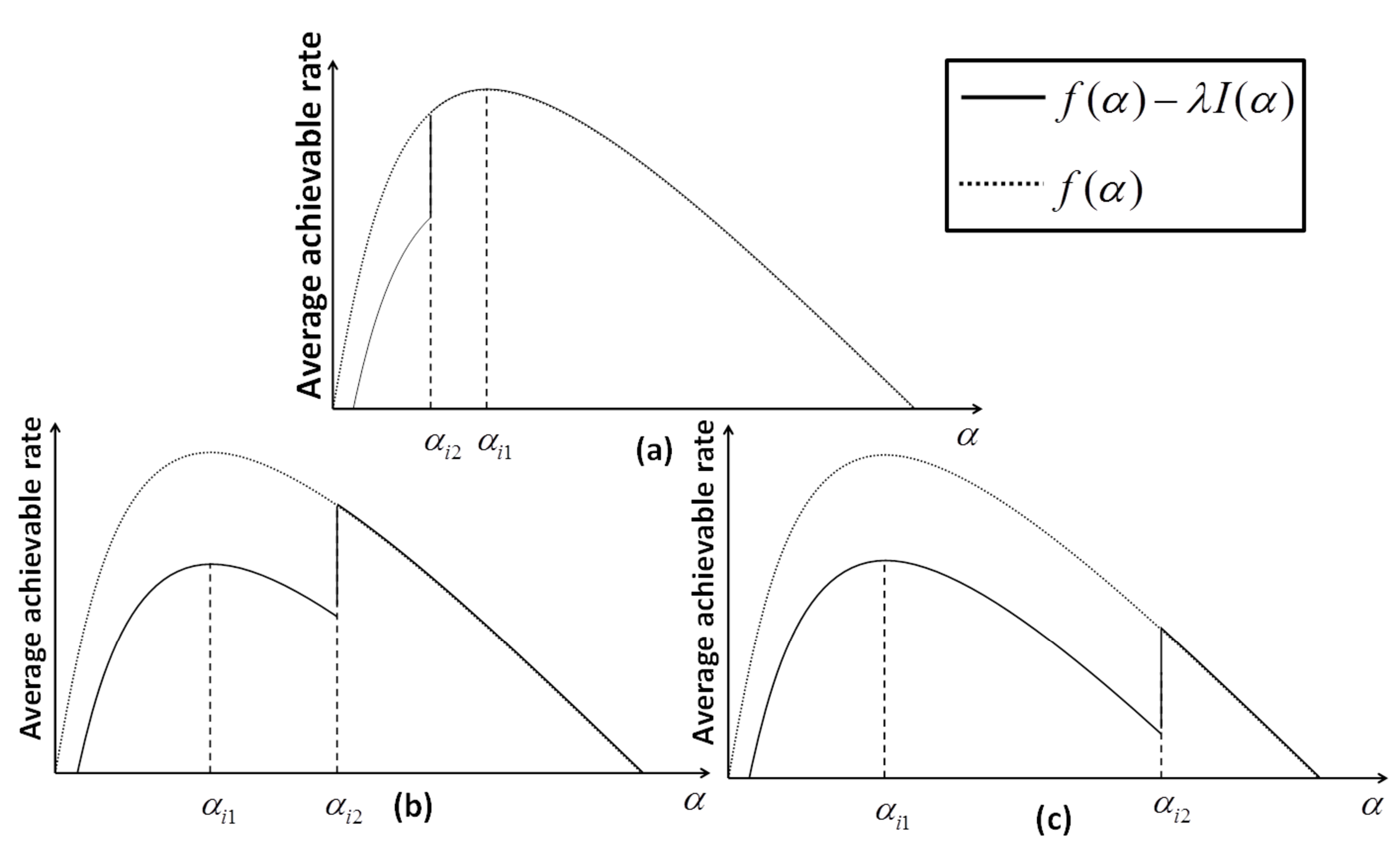}
\caption{Illustration of the optimization problem}
\label{fig:2}
\end{figure}

Note that both $\alpha_{i1}$ and $\alpha_{i2}$ satisfy the second constraint of (\ref{eq:6}) and (\ref{eq:7}). Therefore, (\ref{eq:12})-(\ref{eq:13}) is the optimal solution of our optimization problems \textbf{P1} and \textbf{P2}. Based on (\ref{eq:12}) and (\ref{eq:13}), we can distinguish between the following cases (Fig.~\ref{fig:2}):

\textit{Case 1:} $\alpha _{i1}^{} \ge \alpha _{i2}^{}$. In this case, the optimal value $\alpha_{i1}$ satisfies the interference constraints towards the $R_P$, (\ref{eq:2}), and at this value of $\alpha_{i1}$, the indicator function $I(\alpha_{i1})$ is zero, in which case $R_{i1}^{} = f(\alpha _{i1}^{})$. Hence, the CR system is capable to achieve the maximum possible rate (Fig.~\ref{fig:2}a).

\textit{Case 2:} $\alpha _{i1}^{} < \alpha _{i2}^{}$. In this event, there are two sub-cases depending on the values of  $\alpha_{i1}$ and $\alpha_{i2}$. More specifically, if $R_{i1}^{} < {R_{i2}}$ than the optimal value of the time resource allocation should be equal to  $\alpha_{i2}$ (Fig.~\ref{fig:2}b). If $R_{i1}^{} \geq {R_{i2}}$, than the optimal value $\alpha_{i1}$ maximizes the achievable rate of the penalty introduced by the $\lambda$ parameter (Fig.~\ref{fig:2}c).

\section{Average achievable rates}

This section analyzes the average achievable rates of the considered CR system with RF energy harvesting. The rates are determined in function of the average harvested energy, the channel gain, the outage probability threshold and the PUI threshold.  For this purpose, we use Monte Carlo simulations. The simulation setup is presented in Table~\ref{tab:1}.

\begin{table}[ht]
\caption{Simulation setup}
\label{tab:1}
\centering
\resizebox{0.4\textwidth}{!}{\begin{tabular}{|c||c|}
\hline
\textbf{Parameter types} & \textbf{Parameter values}\\
\hline
\textbf{Simulation environment} & Matlab\\
\hline
\textbf{Number of time slots per simulation run ($M$)} & 10000 \\
\hline
\textbf{Channel model} & Rayleigh fading \\
\hline
\textbf{PU outage probability ($\varepsilon$)} & from 0.01 to 0.99 \\
\hline
\textbf{Noise power ($\sigma^2$)} & -90dBm \\
\hline
\end{tabular}}
\end{table}

For the various curves, we consider realistic values of the channel gain averages $\mu_x$, $\mu_y$, $\mu_z$, $\mu_g$ (from $10^{-10}$ to $10^{-2}$) and PUI thresholds $\gamma_{th}$ (from -90dBm to -30dBm). The average achievable rate is calculated as:

\begin{equation}\label{eq:14}
\bar R = \frac{1}{M}\sum\limits_{i = 1}^M {{R_i}},
\end{equation}

\noindent where $R_i$ is given by (\ref{eq:4}).

\begin{figure}[!t]
\centering
\includegraphics [width=3.5in]{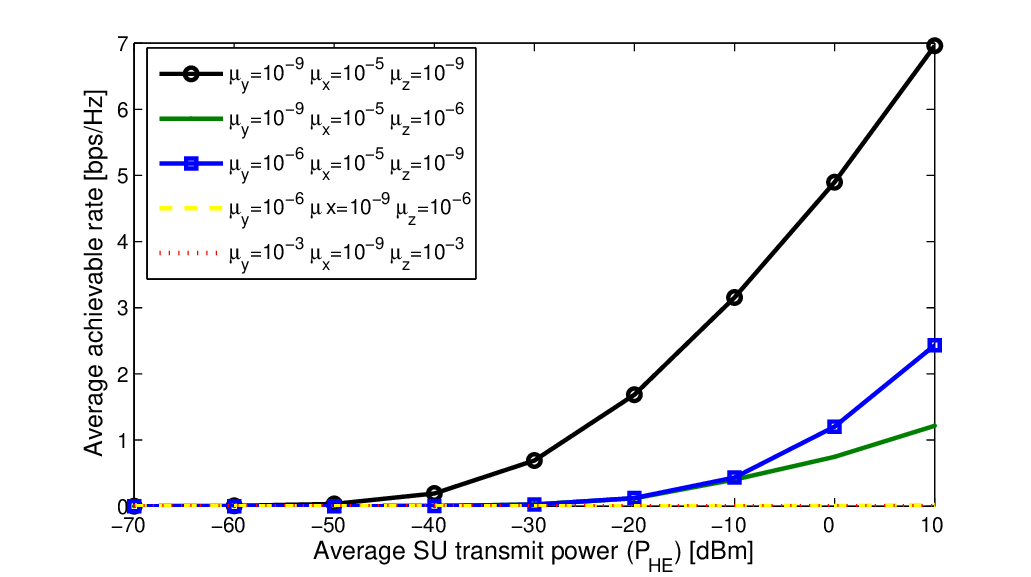}
\caption{Average achievable rate in dependance of the average harvested PU power  (PUI threshold = -90 dBm, Outage probability threshold = 1\%, PU transmit power $P_T$ = 30dBW)}
\label{fig:3}
\end{figure}

Fig.~\ref{fig:3} presents the average achievable rate of the CR system in function of the average SU transmit power ($P_{HE}$), expressed in dBm. The average SU transmit power reflects the actual energy harvested from $T_P$, and is calculated as:

\begin{equation}\label{eq:15}
{P_{HE}} = \mathbb{E}\left[ {\frac{{1 - {\alpha _i}}}{{{\alpha _i}}}{g_i}{P_T}} \right] = \frac{1}{M}\sum\limits_{i = 1}^M {\frac{{1 - {\alpha _i}}}{{{\alpha _i}}}{g_i}{P_T}}.
\end{equation}

The different curves are for the various combinations of $\mu_x$, $\mu_y$ and $\mu_z$. In the best case scenario, when a) the PU and SU transmitters are close to each other, and also b) both receivers experience small amount of interference (i.e. substantially far apart or/weak interference channels $y_i$ and $z_i$), then the CR system achieves practically useful information rates (black circles). The average achievable rate of the SU system decreases significantly when the receivers experience higher interference margins (blue rectangles and doted green lines). More specifically, the CR system achieves better performance when it is subjected to higher interference (blue line with rectangles) instead of causing high interference to the PU receiver (doted green line). This behavior is due to the nature of the underlay spectrum sharing method, which allows the SU to interfere with the PU system in only a fraction of the time ($\varepsilon = 1\%$ in this case). This means that the energy harvesting CR system is more robust and favors scenarios where the pathloss between the SU transmitter and PU receiver is higher compared to the pathloss on the channel between the PU transmitter and the SU receiver.

In the worst case scenario, the harvested energy is small (the PU and SU transmitters are far apart) and the interference channels $y_i$ and $z_i$ experience low path-loss, in which case, the secondary system achieves negligible information rates. This is an expected result, because for very low (i.e. minimal) amounts of harvested energy, the PU has a minimal budget of transmit power. Additionally, because of the low pathloss on the interference channels, the SU receiver suffers from high PU interference, resulting in practically useless information rates (dashed red and yellow lines).

\begin{figure}[!t]
\centering
\includegraphics [width=3.5in]{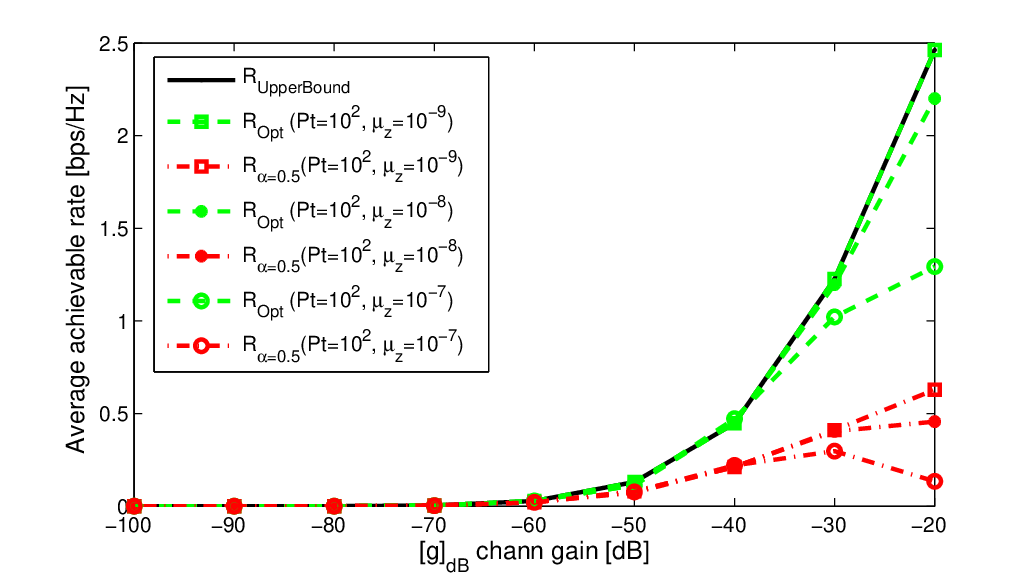}
\caption{Average achievable rate in dependance of the gain on channel $g$ (PUI threshold = -90dBm, Outage probability threshold = 1\%, $\mu_x = 10^{-3}$, $\mu_y = 10^{-7}$)}
\label{fig:4}
\end{figure}

As discussed in the previous figure, the pathloss incurred on channel $z$ crucially affects the average achievable rate. Fig.~\ref{fig:4} depicts the average achievable rate in function of the average channel gain $g$ ($\mu_g$), expressed in dB as $10log10(\mu_g)$, while average channel gain of the channel $z$ ($\mu_z$) appears as a curve parameter. For comparison purposes, the figure also presents the average achievable rate for a system with equal and fixed durations of the EH and IT phases (i.e., $\alpha_i = 0.5, \forall i$). The figure also depicts the theoretical upper bound of the average achievable rate, which is obtained when the primary system constraint, expressed by (\ref{eq:1}), is neglected from the optimization problem (in this case, $\alpha _i^* = \alpha _{i1}^{},\forall i$).

It is evident that the parameter $\mu_z$ significantly affects the performance of the CR system. We also note that our proposed solution significantly outperforms the scenario with $\alpha_i = 0.5$, which justifies the optimal time resource allocation. Moreover, when the channel $z$ is very week (e.g. $\mu_z = 10^{-8}$ or $\mu_z = 10^{-9}$), then the proposed solution achieves rates close to the upper bound for the case $\mu_z = 10^{-8}$, and almost identical to the upper bound for the case $\mu_z = 10^{-9}$. Because of the high path-loss on the interference channel between the SU transmitter and PU receiver, the CR system can exploit the full potential of the harvested energy without causing harmful interference to the PU system while attain information rates close to the upper bound.

\begin{figure}[!t]
\centering
\includegraphics [width=3.5in]{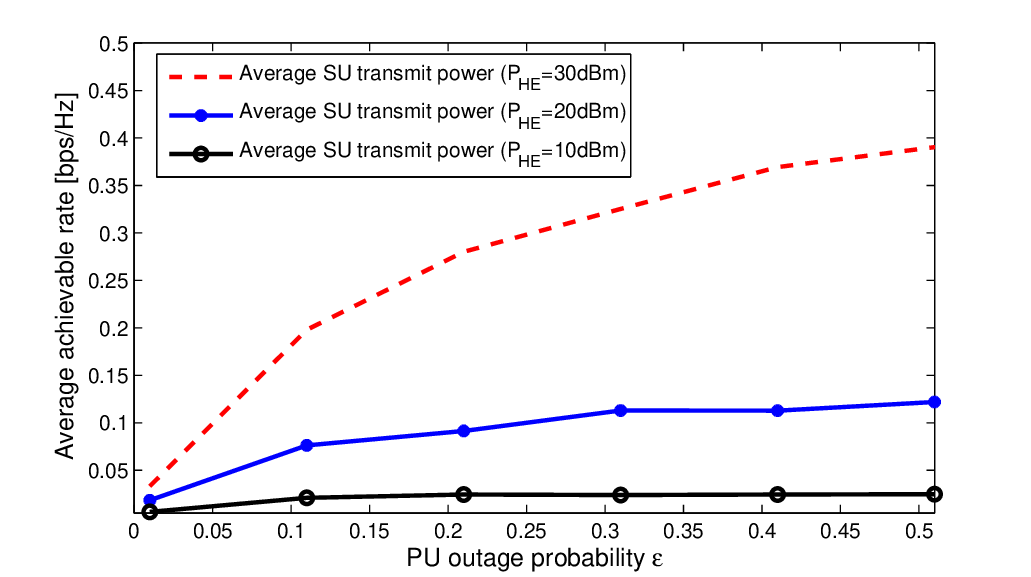}
\caption{Average achievable rate in dependance of the outage probability threshold (PUI threshold = -90dBm , $\mu_y=10^{-7}$, $\mu_x=10^{-5}$, $\mu_z=10^{-6}$)}
\label{fig:5}
\end{figure}

Fig.~\ref{fig:5} depicts the performance of the CR system in function of the PU outage probability threshold ($\varepsilon$). We notice that the average achievable rate of the considered CR system depends on $\varepsilon$ and also on the amount of harvested RF energy. Intuitively, the average achievable rate of the CR system increases when the secondary system causes more frequent outages to the PU system.

\begin{figure}[!t]
\centering
\includegraphics [width=3.5in]{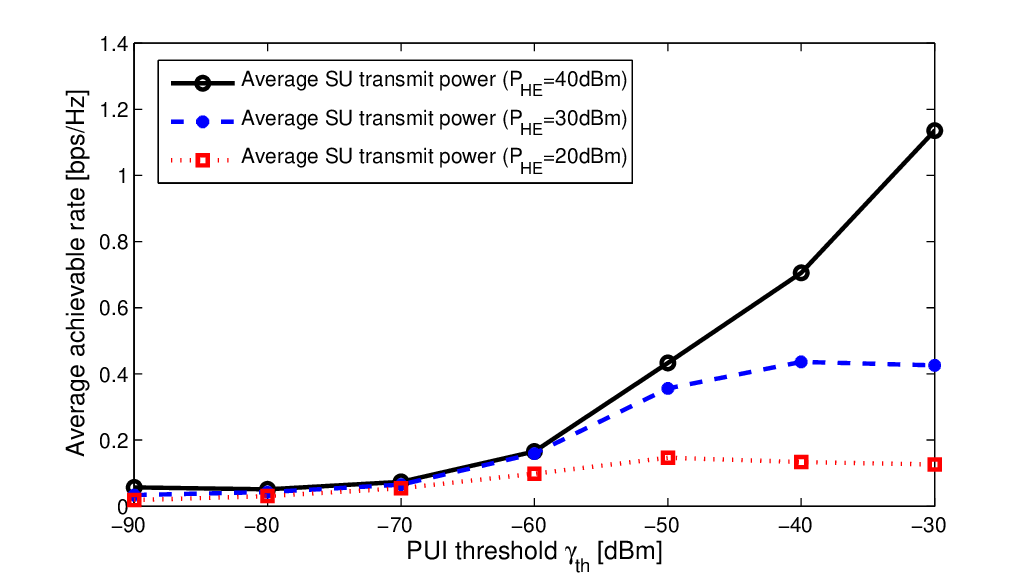}
\caption{Average achievable rate in dependance of the PUI threshold (PU outage probability = 1\%, $\mu_y=10^{-7}$, $\mu_x=10^{-5}$, $\mu_z=10^{-6}$)}
\label{fig:6}
\end{figure}

Fig.~\ref{fig:6} depicts the average achievable rate in function of the PUI threshold ($\gamma_{th}$), expressed in dBm, for different values of the average SU transmit powers ($P_{HE}$). We notice that higher rates are achieved for higher amounts of harvested energy. Additionally, the average achievable rate increases when the PU interference constraint is relaxed (i.e. with increasing $\gamma_{th}$). However, the average achievable rate begins to saturate after some PUI threshold, which in turn heavily depends on the amount of harvested energy. Increasing $P_{HE}$, the saturation point moves towards the higher PUI thresholds. These saturation points can play crucial role when dimensioning and optimizing the CR network. For example, the presented system model and optimization can be easily applied to the scenario of cognitive D2D communication system, where the SU is the D2D system, while the PU system is a mobile cellular system (consisting of a base and mobile stations). In this case it is vital for the D2D devices to achieve the highest possible information rate while considering the amount of harmful generated imposed on the PU (cellular) system. The saturation points of Fig.~\ref{fig:6} gives an idea how to optimize the overall system by achieving the highest information rate in the SU system and low interference in the PU system.

\section{Conclusion}

This work studies an underlay CR system with RF energy harvesting, and determines the optimal time resource allocation of the energy harvesting and information transmission phases. The proposed optimal solution is an online solution that maximizes the average achievable rate of the CR system while limiting the harmful interference from the CR system to the PU system. The simulation results show that the CR system can attain practically useful throughput when subjected to the influence of a powerful primary transmitter. The solution always outperforms the case of fixed time allocation and in some cases comes very close to the information rates that can be achieved when the protection criterion constraint of the primary system is not considered. The proposed system and its optimal solution is applicable to future wireless networks, such as the cognitive D2D communication systems.

\vspace{-2mm}
\section*{Acknowledgment}
This work is supported in part by the Public Diplomacy Division of NATO in the framework of "Science for Peace" through the SfP-984409 "Optimization and Rational Use of Wireless Communication Bands (ORCA)" project. This work is supported in part by the reintegration grant of the Alexander von Humboldt foundation.


\end{document}